\documentclass[showpacs,showkeys]{revtex4}%
\usepackage{amssymb}
\usepackage{amsmath}
\usepackage{amsfonts}
\usepackage{graphicx}
\usepackage{ulem} 
\usepackage{cancel}
\usepackage{epstopdf}

\providecommand{\abs}[1]{\left| #1 \right|} %
\providecommand{\av}[1]{\left\langle #1 \right\rangle} %
\providecommand{\rbr}[1]{\left( #1 \right)}%
\providecommand{\sqbr}[1]{\left[ #1 \right]} %
\def\ra{\rightarrow}

\def\kB{k_{\text{\tiny{B}}}}

\begin{document}

\title[ ]{The Canonical Distribution without Thermodynamic Limit}
\author{$^{1}$Thomas Oikonomou}
\email{thomas.oikonomou@nu.edu.kz}
\author{$^{2}$G. Baris Bagci}
\affiliation{$^{1}$Department of Physics, School of Science and
Technology, Nazarbayev University, 53 Kabanbay Batyr Ave., Astana
010000, Kazakhstan}
\affiliation{$^{2}$Department of Materials Science and Nanotechnology Engineering, TOBB University of Economics and Technology, 06560 Ankara, Turkey}

\begin{abstract}
We derive the continuous canonical distribution only by requiring the extensivity of the mean energy and the multiplicative probabilistic composition rule.
The derivation is independent of the thermodynamic limit and
moreover it does not use the usual equal a priori probability
postulate.
We numerically demonstrate the implications of our derivation for the free and oscillating molecules.
\end{abstract}

\eid{ }
\date{\today }
\startpage{1}
\endpage{1}
\maketitle

\section{Introduction}
%
The concept of statistical equilibrium thermodynamics has been
developed to describe the macroscopic  behaviour of physical
systems in terms of their microscopic structure, namely the
dynamic of their constituent elements such as particles and
molecules.
A pivotal issue of this approach is  the determination of the
energy probability distribution $P(E_r)$ of the system under consideration.
For this, one first needs to determine the probability $\mathcal{P}_j$ that, at any time $t$, the system is to be found in a state $j$ characterized by the energy value $E_j$.
Then, factorizing the former states into groups of states with the same energy levels $E_r$, $j(\text{states})\rightarrow r(\text{levels})$, one obtains the desired energy distribution $P(E_r)=\Omega(E_r)\mathcal{P}_{j\rightarrow r}$, where $\Omega(E_r)$ is the system's degeneracy number of the $r$th energy level.
According to  the fundamental equiprobability postulate of statistical mechanics, all the accessible energy states occur equally likely at thermal equilibrium, so that $\mathcal{P}_j$ is given its classical definition as a probability.

Considering then a system at canonical thermal equilibrium, that is a system of variable energy $E_j$ due to its contact with an $N'$-molecule heat bath,  the probability $\mathcal{P}_j$ is computed as proportional to the microstates $\Omega'(E'_j)$ (equiprobability postulate) of the heat bath, where $E_{\text{tot}}$ denotes the constant total energy of the system plus the thermal bath, i.e., $E_{\text{tot}}=E'_j + E_j$.
Then, considering the heat bath in the thermodynamic limit, $N'\rightarrow\infty$, so that the energy levels $E'_j$ are a continuum, $E'_j\rightarrow E'$ and assuming further that $E'$  is overwhelmingly larger than the energy $E_j$ of the system, thus satisfying the condition $E'\approx E_{\text{tot}}$, one performs  a Taylor expansion of $\Omega'(E')$ around $E'\rightarrow E_\mathrm{tot}$ to obtain \cite{Pathria}
\begin{equation}\label{Intro:01a}
\mathcal{P}_j\propto \Omega'(E'_j)\propto e^{-\beta' E_j}\,,
\end{equation}
where $\beta':= \frac{\partial \ln \Omega'(E_\mathrm{tot})}{\partial E_\mathrm{tot}}$. 
The exponential term in Eq. (\ref{Intro:01a}) is called Boltzmann factor.
%
%
Then, the canonical energy distribution of the system is determined as
\begin{equation}\label{Intro:02}
P(E_r)=\Omega(E_r)\mathcal{P}_{j\rightarrow r}=\frac{\Omega(E_r)\, e^{-\beta' E_r}}{\sum_r \Omega(E_r)\, e^{-\beta' E_r}}\,.
\end{equation}
%
%
Considering the system in the thermodynamic limit as well, $N\rightarrow\infty$, so that $E_r\rightarrow E$ and $\Omega(E_r)$ can be expressed as $\Omega(E)\mathrm{d}E$, where $\Omega(E)$ is now the density of states, Eq. \eqref{Intro:02} takes its continuous form as
\begin{equation}\label{Intro:03}
P(E)=\frac{\Omega(E)\, e^{-\beta' E}\mathrm{d}E}{\int \Omega(E)\, e^{-\beta' E}\mathrm{d}E}\,.
\end{equation}
We stress though that the passage from Eq. \eqref{Intro:02} to Eq. \eqref{Intro:03} is not strictly derived, but it is justified as the most natural choice \cite{Chandler}.

%
%
%

As we have seen above, four assumptions have been invoked for the derivation of Eq. (\ref{Intro:03}), i.e., the equiprobability postulate as the thermal equilibrium condition, a system of negligible energy compared to the energy of the heat bath, and the thermodynamic limit of both the heat bath and the system in order to obtain continuous energies and being able to apply the calculus. 
It is thus scientifically an intriguing question to explore whether there is a way to derive the energy distribution in the canonical case, by minimizing or even if possible eliminating the preceding assumptions and how this would affect the final results.
In an effort to answer this question, in this work, we follow a novel approach to derive the energy distribution of a system composed of $N$ identical molecules at the canonical equilibrium. The cornerstone of our approach is, instead of equal probabilities, to use the internal energy extensivity property, i.e.,  the  proportionality to $N$, as the thermal equilibrium condition.
Interestingly enough then, non of the four assumptions are needed for the derivation of the canonical distribution within this approach. The results, as expected, are shown to be more general than the textbook ones,  providing new perspectives within statistical thermodynamics, which we indent to explore closer in the future.

For this purpose, we first define in Section \ref{Ensemble} the canonical ensemble of discrete energy states through a minimum number of statistical mechanical conditions (excluding thereby the equiprobability postulate), showing that it satisfies indeed the energy extensivity. 
However, the related discrete energy probability distribution $P_r$, is not to be considered at this stage as the canonical distribution. Its structural generality is to be reduced by requesting the validity of the equilibrium condition in the continuous limit as well.
This is done  in Section \ref{CED}, where we extend the discussion to the continuous case.
The obtained continuous distribution $P(E)$ is now the canonical one and by discretizing it we determine $P_r$ for the discrete energies levels.
Our results show that the currently derived canonical distribution contains the Boltzmann factor $e^{-\beta E_j}$, as in Eq. (\ref{Intro:01a}), yet its origin is different and the energy factor $\beta$, in contrast to $\beta'$, is not subjected a specific statistical structure.
In Section \ref{numerics}, we present some numerical results to support our findings.
Finally, discussion and remarks are presented in the conclusions.

\section{Discrete Canonical Ensemble}\label{Ensemble}
%
In this section we will introduce the discrete energy ensemble describing a system being at canonical equilibrium.
We consider therefore a closed system composed of $N$ molecules plus the reservoir. We  denote the  sample space of all possible mutually exclusive discrete energies $\varepsilon_i^{(\nu)}$ of the $\nu$th molecule by $\omega_\nu\equiv \{\varepsilon_i^{(\nu)}\}_{i=1, \ldots, \alpha\in\mathbb{N}}$ with $\nu=1,\ldots,N$.
Then, the probability of finding the $\nu$th molecule with energy $\varepsilon_i^{(\nu)}$ is denoted as $p_\nu( \varepsilon_i^{(\nu)})$. 
It satisfies the following normalization condition
%
\begin{equation}\label{NormConstr1}%
p_\nu(\omega_\nu)
=\sum_{i=1}^{\alpha}p_\nu(\varepsilon^{(\nu)}_i)=1
\qquad \text{and}\qquad
0<p_\nu(\varepsilon_i^{(\nu)})<1\,,
\end{equation}
%
in each sample space $\omega_\nu$.
The boundary  values $0$ and $1$ in the double  inequality in Eq. \eqref{NormConstr1} are excluded since
$p_\nu(\varepsilon_i^{(\nu)})=1$ would imply the existence of a
unique  energy value $\varepsilon_i^{(\nu)}$ and
$p_\nu(\varepsilon_i^{(\nu)})=0$ would imply that
$\varepsilon_i^{(\nu)}$ is not a constituent element of
$\omega_\nu$.
In other words,  $\omega_\nu$ contains all the accessible energy values and only them.
For the sake of  simplicity, we assume in what follows that the
molecules are identical, i.e.,
$(\omega_\nu,p_\nu)=(\omega_{\nu'},p_{\nu'})=(\omega,p)$. The
index $\nu$ will be used, when necessary, only for heuristic
reasons.
Then, the canonical ensemble of the total $N$-molecule system is defined by the following three conditions:
\begin{itemize}
\item[C1.] The sample space $\mathcal{W}$ of the energy states
$\{\mathcal{A}_j\}_{j=1,\ldots,W\in\mathbb{N}}$ is determined by
a the tensor product $\otimes_{\wedge}$ of the sets $\omega_\nu$
over the conjunction operator $\bigwedge$ \cite{Comment1},
%
\begin{align*}
\mathcal{W} &=
\omega_1\otimes_{\wedge}\omega_2
    \otimes_{\wedge}\cdots \otimes_{\wedge}\omega_{N-1}\otimes_{\wedge}\omega_N\\
&=\Bigg\{
\bigwedge_{\nu=1}^{N}\varepsilon_{1}^{(\nu)},
\bigwedge_{\nu=1}^{N-1}\varepsilon_{1}^{(\nu)}\bigwedge \varepsilon_{2}^{(N)}, \ldots,
\bigwedge_{\nu=1}^{N-1}\varepsilon_{\alpha}^{(\nu)}\bigwedge \varepsilon_{\alpha-1}^{(N)},
\bigwedge_{\nu=1}^{N}\varepsilon_{\alpha}^{(\nu)}\Bigg\}\\
&=:\big\{\mathcal{A}_1,\mathcal{A}_2,\ldots,\mathcal{A}_{W-1},\mathcal{A}_{W}\big\}\,,
\end{align*}
%
where $W\equiv W(\alpha,N)$ is the cardinality of   $\mathcal{W}$ computed as
\begin{equation}\label{ConfFun-ind-a}
W(\alpha,N)=\underbrace{\alpha\times\alpha\times\cdots\times\alpha\times\alpha}_{N\; \text{times}}=\alpha^N\,.
\end{equation}
\item[C2.] The probability of occurrence for the $j$th state is described by  the multiplicative  composition rule, e.g., $\mathcal{P}(\bigwedge_{\nu=1}^{N} \varepsilon^{(\nu)}_i) = \prod_{\nu=1}^{N} p_\nu(\varepsilon^{(\nu)}_i)$. $\mathcal{P}$  has  to satisfy the analogous relations to Eq. (\ref{NormConstr1}), namely
\begin{equation}\label{JPNorm1}
\mathcal{P}(\mathcal{W})= \sum_{j=1}^{W} \mathcal{P}(\mathcal{A}_j)=1
\qquad \text{and}\qquad
0<\mathcal{P}(\mathcal{A}_j)<1\,,
\end{equation}
It is worth stressing a very misused issue in literature, namely that if the energy levels $\varepsilon^{(\nu)}_i$ are statistically independent then the multiplicative composition rule holds, yet not vice versa \cite{example}. Indeed, the application of the former composition rule may describe statistically dependent $\varepsilon_i$'s as well.
\item[C3.] The energy $E_j\equiv E(\mathcal{A}_j)$ of the $j$th  state is additive: if $n_{ij}\equiv n_j(\varepsilon_i) $ is the frequency of the energy value $\varepsilon_i$  within the state $\mathcal{A}_j$, then for any $j$, the energy $E_j$ is given as
%
\begin{equation}\label{StateEnergy_j}
E_j=\sum_{i=1}^{\alpha}n_{ij}\varepsilon_i
\qquad\qquad\text{with}\qquad\qquad
\sum_{i=1}^{\alpha}n_{ij}=N\,.
\end{equation}
%
\end{itemize}
The conjunction  sign $\bigwedge$ in a configuration, e.g.,
$\bigwedge_{\nu=1}^{N}\varepsilon^{(\nu)}_i$, simply implies:
$\varepsilon_i^{(1)}\;\text{and}\;\varepsilon_i^{(2)}\;\text{and}\cdots\text{and}
\;\varepsilon_i^{(N)}$.
The probability of occurrence for the $j$th state, due to the
condition C2, is formed as
%
\begin{align}\label{Prob2}
\begin{aligned}
\mathcal{P} (\mathcal{A}_1) = \mathcal{P}\rbr{\bigwedge_{\nu=1}^{N}\varepsilon_{1}^{(\nu)}}
&=\prod_{\nu=1}^{N}p_\nu\rbr{\varepsilon_1^{(\nu)}},\\
\mathcal{P}(\mathcal{A}_2) = \mathcal{P}\rbr{\bigwedge_{\nu=1}^{N-1}\varepsilon_{1}^{(\nu)} \bigwedge \varepsilon_{2}^{(N)}}
&=\prod_{\nu=1}^{N-1}p_\nu\rbr{\varepsilon_1^{(\nu)}}p_N\rbr{\varepsilon_2^{(N)}},\\
&\;\;\vdots\\
\mathcal{P}(\mathcal{A}_{W-1}) = \mathcal{P}\rbr{\bigwedge_{\nu=1}^{N-1}\varepsilon_{\alpha}^{(\nu)} \bigwedge\varepsilon_{\alpha-1}^{(N)}}
&=\prod_{\nu=1}^{N-1}p_\nu\rbr{\varepsilon_\alpha^{(\nu)}}
    p_N\rbr{\varepsilon_{\alpha-1}^{(N)}},\\
\mathcal{P}(\mathcal{A}_{W}) = \mathcal{P} \rbr{\bigwedge_{\nu=1}^{N}\varepsilon_{\alpha}^{(\nu)}}
&=\prod_{\nu=1}^{N}p_\nu\rbr{\varepsilon_\alpha^{(\nu)}}.
\end{aligned}
\end{align}
%
For identical molecules,   Eq. \eqref{Prob2} can be written in the compact form
%
\begin{equation}\label{JointProb-jNotation}
\mathcal{P}_{j}=\prod_{i=1}^{\alpha}p_i^{n_{ij}}\,,
\end{equation}
%
where $\mathcal{P}_j\equiv \mathcal{P}(\mathcal{A}_j)$ and $p_{i}\equiv p(\varepsilon_i)$.
By the multinomial theorem (see Appendix A for more details), we obtain
\begin{equation*}\label{MC_relation_1}
\sum_{j=1}^{W} \mathcal{P}_j = \rbr{\sum_{i=1}^{\alpha}p_i}^N \,,
\end{equation*}
which yields
\begin{equation}\label{MC_relation_2}
\sum_{j=1}^{W} \mathcal{P}_j=1 \,,
\end{equation}
as a result of Eq. \eqref{NormConstr1}. Moreover, applying the operator $p_k \frac{\partial}{\partial p_k}$ (see Appendix A), we obtain the following general relation valid within the  sample space $\mathcal{W}$
\begin{equation*}\label{NewEq2_1}
\sum_{j=1}^{W} \mathcal{P}_j \,n_{ij} = N\,p_i\rbr{\sum_{i=1}^{\alpha}p_i}^{N-1}
\end{equation*}
which yields
\begin{equation}\label{NewEq2_2}
\sum_{j=1}^{W} \mathcal{P}_j\,n_{ij}=N\,p_i
\end{equation}
again as a result of Eq. \eqref{NormConstr1}.
As can be seen here, the probability measure in Eq. \eqref{JointProb-jNotation}   satisfies indeed the normalization condition as a consequence of Eq. \eqref{NormConstr1} and the conditions C1-C2 for any energy value $\varepsilon_i$ and any arbitrary  structure of  $p_i$.

Having determined the structure of the probability measure $\mathcal{P}_j$ yielding the likelihood of the occurrence of the $j$th state in Eq. \eqref{JointProb-jNotation}, we may now consider the likelihood of the occurrence of the states with the same energy.
To this aim, we relabel the $\mathcal{A}_j$ with a new  index $r=1,\ldots, w$, so that each $r$ corresponds to a set of states exhibiting the same energy $E_r$.
Apparently, $E_r$ satisfies Eq. \eqref{StateEnergy_j} for $j\ra r$.
Then, by virtue of  Eq. \eqref{JointProb-jNotation}, we  determine the probability with which  a state occurs with the energy $E_r$ as
\begin{equation}\label{ESProbDF}
P_r= \Omega(E_r) \mathcal{P}_{j\rightarrow r}=\Omega(E_r) \prod_{i=1}^{\alpha}p_i^{n_{ir}}\,,
\end{equation}
where $P_r\equiv P(E_r)$ and $\Omega(E_r)$  is the degeneracy number of the $r$th energy value of the system.
In the general case of a nonlinear dependence  of $\varepsilon_i$ on $i$, $\Omega(E_r)$ is given by the multinomial coefficient with $r_{\max}\equiv w\equiv w(\alpha,N)=(N+\alpha-1)!/\big[(\alpha-1)!N!\big]$, so that $\sum_{r=1}^{w}\Omega(E_r)=W=\alpha^N$ (see Appendix A).
Apparently, the energy probability distribution  in Eq. \eqref{ESProbDF} is normalized within $\mathcal{W}$, since $\sum_{r=1}^{w}P_r=\sum_{j=1}^{W} \mathcal{P}_j$.

By virtue of Eq. \eqref{NewEq2_2} then, we can show that the mean energy $\av{E}$ of the ensemble $\mathcal{W}$ is proportional to the number of molecules as
%
\begin{equation}\label{D_AvEn}
\av{E} 
= \sum_{r=1}^{w}P_r\,E_r 
= \sum_{j=1}^{W} \mathcal{P}_j\,E_j
= \sum_{j=1}^{W} \sum_{i=1}^{\alpha} \mathcal{P}_j\,n_{ij}\, \varepsilon_i
= \sum_{i=1}^{\alpha}\bigg[\sum_{j=1}^{W} \mathcal{P}_j \, n_{ij}\bigg]\, \varepsilon_i  
= N\sum_{i=1}^{\alpha} p_i\, \varepsilon_i=N\av{\varepsilon}\,,
\end{equation}
%
where $\av{\varepsilon}$ is the mean energy of a single molecule.
Identifying $\av{E}$ with the internal energy of the system, we see that the ensemble $\mathcal{W}$ under the conditions C1-C3 satisfies indeed the thermal equilibrium condition, i.e., the energy extensivity, justifying the denomination of the discrete canonical ensemble. 
It is worth remarking, that Eq. \eqref{D_AvEn} is valid only as long as $N$ and $\alpha$ are finite, warranting  the finiteness required to interchange the order of the summation. %

\section{Continuous Canonical Ensemble}\label{CED}
%
In this section, our aim is to derive the continuous version of Eq. \eqref{ESProbDF} subject the maintenance of the energy extensivity.
For this, we first need to find a passage to transit from discrete to continuous energies, $E_r\ra E$. There are two options for this, the textbook one, i.e., the thermodynamic limit of the system $N\rightarrow\infty$, or $\alpha \rightarrow \infty$. Considering  the discrete energy expression in Eq. (\ref{StateEnergy_j}) for $j\ra r$, i.e. $E_r=\sum_{i=1}^{\alpha} n_{ir} \varepsilon_i$, we can see that the only quantity exhibiting a dependence on $N$ is the frequency $n_{ir}$ of finding the energy value $\varepsilon_i$ within the energy state $\mathcal{A}_r$. However, irrespective of the number of the constituent molecules of the system the image of $n_{ir}$ is always an integer number. Therefore, considering the limit $N\ra\infty$, the energy values $E_r$ do not become continuous.
Thus, the only option left for the system energy to become continuous is to assume that the discrete energy values $\varepsilon_i \in[a,b] \subseteq \mathbb{R}_+$ becomes more and more numerous for increasing $\alpha$ \cite{Jaynes}. In this way, for $\alpha\ra\infty$ these discrete energies become continuous, $\varepsilon_i \rightarrow \varepsilon$ ($E_r\rightarrow E$), albeit in the same range $[a,b]$ ($[Na,Nb]$). Accordingly, in the former limit the difference between two successive energy values tends to zero, $\Delta \varepsilon_i \rightarrow 0$ ($\Delta E_r\rightarrow0$).

Following Jaynes \cite{Jaynes}, we then consider the respective discrete energy probability distribution as
\begin{equation}\label{Sec3:01}
p(\varepsilon_i)
=\frac{f(\varepsilon_i)\Delta \varepsilon^*_i}{z_d}\,,
\qquad\qquad
z_d:=\sum_{i=1}^{\alpha} f(\varepsilon_i)  \Delta\varepsilon^*_i\,,
\end{equation}
where $\Delta \varepsilon^*_i:= \varepsilon^*_{i+1} - \varepsilon^*_{i}$ with $\varepsilon^*_1\equiv \varepsilon_1$, $\varepsilon^*_{\alpha+1}\equiv \varepsilon_\alpha$ and $\varepsilon_{1<i\leq \alpha}^* \in[\varepsilon_{i-1},\varepsilon_{i}]$.
In the limit $\alpha\rightarrow\infty$ we have $\Delta\varepsilon_i\rightarrow0$ and thus $\Delta \varepsilon_i^*\rightarrow0$, so that the measure
$p(\varepsilon_i)$ takes its continuous form as
\begin{equation}\label{Sec3:02}
p(\varepsilon)=\frac{f(\varepsilon)\mathrm{d}\varepsilon}{z}\,,
\qquad\qquad
z=\int_a^b  f(\varepsilon)  \mathrm{d}\varepsilon\,,
\end{equation}
where $\lim_{\alpha\rightarrow\infty} f(\varepsilon_i) \rightarrow f(\varepsilon)$ is assumed.
The substitution of Eq. \eqref{Sec3:01} into Eq. \eqref{ESProbDF}
yields
\begin{equation}\label{DiscExpr1a}
P(E_r)=\frac{F(E_r)\,\Delta E^*_r}{Z_d}\,
\end{equation}
with $Z_d:=z_d^N$  and
\begin{equation}\label{DiscExpr1b}
F(E_r) := \frac{\Omega(E_r)}{\Delta
E^*_r}\exp\rbr{\sum_{i=1}^{\alpha}n_{ir}
\ln\big(f(\varepsilon_i)\Delta\varepsilon^*_i\big)}\,,
\end{equation}
where $\Delta E^*_r:=E^*_{r+1}-E^*_r$, $E^*_1\equiv E_1$, $E^*_{w+1} \equiv E_w$, $E_{1<r\leq w}^*\in[E_{w-1},E_{w}]$ and $E_r\in[Na,Nb]\subseteq \mathbb{R}$.
$F(E_r)$ converges to a continuous function $F(E)$ for
$\alpha\rightarrow\infty$  i.e.,  $F(E):=
\lim_{\alpha\ra\infty}F(E_r)$ (see Appendix B), so that the energy probability distribution in Eq. \eqref{DiscExpr1a} becomes continuous as well
\begin{equation}\label{ContExp1a}
P(E)=\frac{F(E)   \mathrm{d}E}{Z} \,,\qquad\qquad
Z=\int_{N a}^{N b} F(E)\,\mathrm{d}E=
\rbr{\int_a^b  f(\varepsilon)\,\mathrm{d}\varepsilon}^N=z^N\,.
\end{equation}

In order to proceed further, we assume that both $Z$ and $z$
depend on an external positive parameter called $\beta$ whose physical
meaning is undetermined for now. Having assumed this dependence,
we take the logarithm of both sides of the equation above, namely
$Z=z^N$, and then take derivative of both sides with respect to
$\beta$. This yields

\begin{equation}
\frac{1}{Z}\int_{Na}^{Nb} \frac{\partial F(E,\beta)}{\partial
\beta} \mathrm{d}E= \frac{N}{z}\int_{a}^{b} \frac{\partial
f(\varepsilon,\beta)}{\partial \beta}  \mathrm{d}\varepsilon\,.
\end{equation}
Enforcing the extensivity of the mean energy i.e.,
$\av{E}=N\av{\varepsilon}$ (see Eq. \eqref{D_AvEn} above),  also
in the continuous case, leaves us with the following two distinct
conditions: either one has
\begin{equation}\label{Cond_1}
\frac{\partial F(E,\beta)}{\partial \beta}=E\,F(E,\beta)
\qquad\text{and}\qquad \frac{\partial
f(\varepsilon,\beta)}{\partial
\beta}=\varepsilon\,f(\varepsilon,\beta)\,.
\end{equation}
or
\begin{equation}\label{Cond_2}
\frac{\partial F(E,\beta)}{\partial \beta}=-E\,F(E,\beta)
\qquad\text{and}\qquad \frac{\partial
f(\varepsilon,\beta)}{\partial
\beta}=-\varepsilon\,f(\varepsilon,\beta)\,.
\end{equation}
In fact, both conditions satisfy the extensivity property
$\av{E}=N\av{\varepsilon}$ in the form
\begin{equation}
\av{E}=\pm\int_{Na}^{Nb} E P(E)=\pm\frac{1}{Z}\int_{Na}^{Nb}E
F(E,\beta)\mathrm{d}E\pm=N \int_{a}^{b} \varepsilon
p(\varepsilon)=\pm\frac{N}{z}\int_{a}^{b} \varepsilon
f(\varepsilon,\beta)\mathrm{d}\varepsilon=N\av{\varepsilon}
\end{equation}
in accordance with Eq. (\ref{ContExp1a}).
The conditions (\ref{Cond_1})-(\ref{Cond_2}) yield
\begin{equation}\label{ContExp1c}
F(E,\beta)=\Phi(E)e^{\pm\beta E}\,,\qquad\qquad
f(\varepsilon,\beta)=\phi(\varepsilon)e^{\pm\beta \varepsilon}\,.
\end{equation}
where the plus sign corresponds to the solution of Eq.
(\ref{Cond_1}) while the minus sign corresponds to the solution of
Eq. (\ref{Cond_2}).
Substituting these two distinct solutions into Eqs.
(\ref{Sec3:02}) and (\ref{ContExp1a}), we obtain
\begin{equation}\label{ContExp1}
p(\varepsilon)=\frac{\phi(\varepsilon)e^{\pm\beta \varepsilon}
\mathrm{d}\varepsilon}{z}\,,\qquad   P(E)=\frac{\Phi(E)
e^{\pm\beta E} \mathrm{d}E}{Z} \,.
\end{equation}
However, since the condition with the plus sign above causes the probability distribution to diverge for high energies ($E\rightarrow\infty$), the relevant sign has to be minus.
Therefore, we finally obtain the continuous version of the canonical distribution as
\begin{equation}\label{FinalContProb1}
p(\varepsilon)=\frac{\phi(\varepsilon)e^{-\beta \varepsilon}
\mathrm{d}\varepsilon}{z}\,,\qquad   P(E)=\frac{\Phi(E) e^{-\beta
E} \mathrm{d}E}{Z} \,,
\end{equation}
with
\begin{equation}\label{FinalContProb2}
Z=\int_{N a}^{N b}\Phi(E) e^{-\beta E}\,\mathrm{d}E=\rbr{\int_a^b
\phi(\varepsilon)
e^{-\beta\varepsilon}\,\mathrm{d}\varepsilon}^N=z^N\,.
\end{equation}
Apparently, the relation between $\Phi$ and $\phi$ is uniquely determined by the finite inverse Laplace transformation \cite{Laplace}
\begin{equation}\label{LTrafo}
\Phi(E)=\frac{1}{2\pi\textrm{i}}
\int_{\beta'-\mathrm{i}\infty}^{\beta'+\mathrm{i}\infty}
\sqbr{\int_a^b \phi(\varepsilon)e^{-\beta
\varepsilon}\mathrm{d}\varepsilon}^N e^{\beta E}
\mathrm{d}\beta\,.
\end{equation}
The probability $P(E)$ reduces to $p(\varepsilon)$ for
$N=1$ as expected. However, note that $P(E)$
is valid for any number of molecules (see also Ref. \cite{OikGBB2013a} for a similar reasoning along the lines of resolving the Gibbs paradox).
%

As we have seen above, in the current approach the canonical energy distribution in Eq. (\ref{FinalContProb1}) is derived based on  two relations, i.e.,  $Z=z^N$ and $\av{E}=N\av{\varepsilon}$. 
The former relation is obtained from the ensemble conditions C1 and C2, and the latter relation is obtained when all  three ensemble conditions C1-C3 are taken into account.
When this is the case, then the Boltzmann factor $e^{-\beta E}$ in the canonical energy distribution emerges.

\section{Derivation of $\Phi(E)$ for $\beta^{-1}-$proportional energies}
We we want to derive the most general expression of the function $\Phi(E)$ for the case where the ensemble energy is inverse proportional to $\beta$. 
Since the mean energy of the ensemble is extensive, we shall only consider a single molecule, thus
\begin{equation}
\av{\varepsilon}=\frac{\lambda}{\beta}\,,
\end{equation}
where $\lambda>0$ is the proportionality constant.
Writing explicitly the averaging formula we obtain
\begin{equation}
\frac{\int_0^\infty \varepsilon\,\phi(\varepsilon)e^{-\beta
\varepsilon}\mathrm{d}\varepsilon}{\int_0^\infty
\phi(\varepsilon) e^{-\beta
\varepsilon}\mathrm{d}\varepsilon}
=\frac{\lambda}{\beta}
\quad\Rightarrow\quad \int_0^\infty
\varepsilon\,\phi(\varepsilon)e^{-\beta
\varepsilon}\mathrm{d}\varepsilon = \frac{1}{\beta}
\int_0^\infty
\lambda\,\phi(\varepsilon)e^{-\beta
\varepsilon}\mathrm{d}\varepsilon
\end{equation}
Partial integration of the l.h.s. yields
\begin{equation}
-\frac{1}{\beta}\varepsilon\,\phi(\varepsilon)e^{-\beta
\varepsilon}\bigg|_0^\infty+\frac{1}{\beta}\int_0^\infty e^{-\beta
\varepsilon}\Big[\phi(\varepsilon)+\varepsilon\phi'(\varepsilon)\Big]\mathrm{d}\varepsilon=\frac{1}{\beta}\int_0^\infty
\lambda\,\phi(\varepsilon)e^{-\beta
\varepsilon}\mathrm{d}\varepsilon
\end{equation}
The first term is equal to zero, assuming a finite contribution of $\phi(0)$. Then, we obtain the following differential equation to
solve
\begin{equation}
\phi(\varepsilon)+\varepsilon\phi'(\varepsilon) = \lambda \phi(\varepsilon)
\end{equation}
yielding
\begin{equation}
\phi(\varepsilon)=c\,\varepsilon^{\lambda-1}\,.
\end{equation}
%
%
Substituting the former result in Eq. (\ref{LTrafo}) we determine the function $\Phi(E)$ for $\beta>0$ as
\begin{equation}
\Phi(E)=\frac{c^N\Gamma^N(\lambda)}{\Gamma(\lambda N)} E^{\lambda N-1}\,.
\end{equation}
Then,   the probabilities $p(\varepsilon)$ and $P(E)$ are computed to be
\begin{equation}\label{Ex:eq02}
p(\varepsilon)=\frac{\beta^{\lambda}}{ \Gamma(\lambda)}
\,\varepsilon^{\lambda-1} e^{-\beta
\varepsilon}\mathrm{d}\varepsilon\,,\qquad
P(E)=\frac{\beta^{\lambda N}}{\Gamma(\lambda N)}
E^{\lambda N-1}e^{-\beta\,E} \mathrm{d}E\,,
\end{equation}
from which we determine the mean energy as well, namely
\begin{equation}\label{Ex:eq03}
\av{E}=\frac{\lambda N}{\beta}=N\av{\varepsilon}\,.
\end{equation}
This is a novel result.
Obviously, $P(E)$ yields $p(\varepsilon)$ for $N=1$. 
Since the energy probability distribution functions $p(\varepsilon)$ and $P(E)$ have to be dimensionless, we read that the parameter $\lambda$ is a merely a number while the factor $\beta$ has the dimension of inverse energy. 
Comparing this result with the kinetic gas theory we identify $\beta=(\kB T)^{-1}$. The value of $\lambda$ depends on the on-molecule potential and the degrees of freedom of a system's molecule.

\section{System-Heat Bath interaction}\label{numerics}
In this section we shall apply numerical analysis to verify our theoretical results. 
Namely, the canonical distribution is valid as long as the system-heat bath interaction preserve the conditions C1-C3, irrespective of the size of the heat bath.

For our purpose we shall consider a system of $N$ free molecules embedded  in an one dimensional  heat bath comprised of $N_\mathrm{HB}$ molecules.
Considering the Langevin dynamic of the free molecules system under consideration with unity mass, the system-heat bath interaction is described by a white noise $\xi$ satisfying the properties $\av{\xi_\nu(t)} =0$, $\av{\xi_\nu ^2(t)}=2\gamma \kB T$,  i.e.,
\begin{equation}\label{LangevinDyn}
\dot{v}_\nu=-\gamma v_\nu  + \xi_\nu(t)\,,
\end{equation}
where $\gamma$ is  the drift parameter, $v_\nu$ and $x_\nu$ are the $\nu$th molecule velocity and position, respectively, and $V$ is the on-molecule potential. $T$ denotes the heat bath's temperature and $\kB$ is the Boltzmann constant. 
From the well known equilibrium expression of the average square velocity obtained from the Langevin dynamic,  we are able to identify the factor $\beta$ with the inverse temperature as $\beta=(\kB T)^{-1}$ and $\lambda=1/2$, so that the energy probability density in Eq. (\ref{Ex:eq02}) reduces to
\begin{equation}\label{EnergyDistr_FreeMol}
\rho_p(\varepsilon)\sim  \varepsilon^{-1/2} e^{-\beta\varepsilon} \,,\qquad
\rho_P(E)\sim E^{N/2-1}e^{-\beta\,E}\,,
\end{equation}
%
%

If the heat bath is of infinite size, then  due to the Central Limit Theorem, the noise is justified to be described by a Gaussian distribution density.
%
%
%
For a finite  heat bath on the other hand,   the white noise is commonly modeled by the Poisson jump-noise \cite{Kopke}
\begin{eqnarray}
\xi(t)=\sum_{k=1}^{n(t)}z_k \delta(t-\tau_k)-\mu \av{z}\,.
\end{eqnarray}
The physical meaning of $\mu$ is the average number of collisions per time interval $\Delta t$. When $\mu\to \infty$ then the Poisson noise recovers the Gaussian noise (infinite many collisions).

\begin{figure}[t]
\begin{center}
    \includegraphics[width=8.1cm,height=6cm]{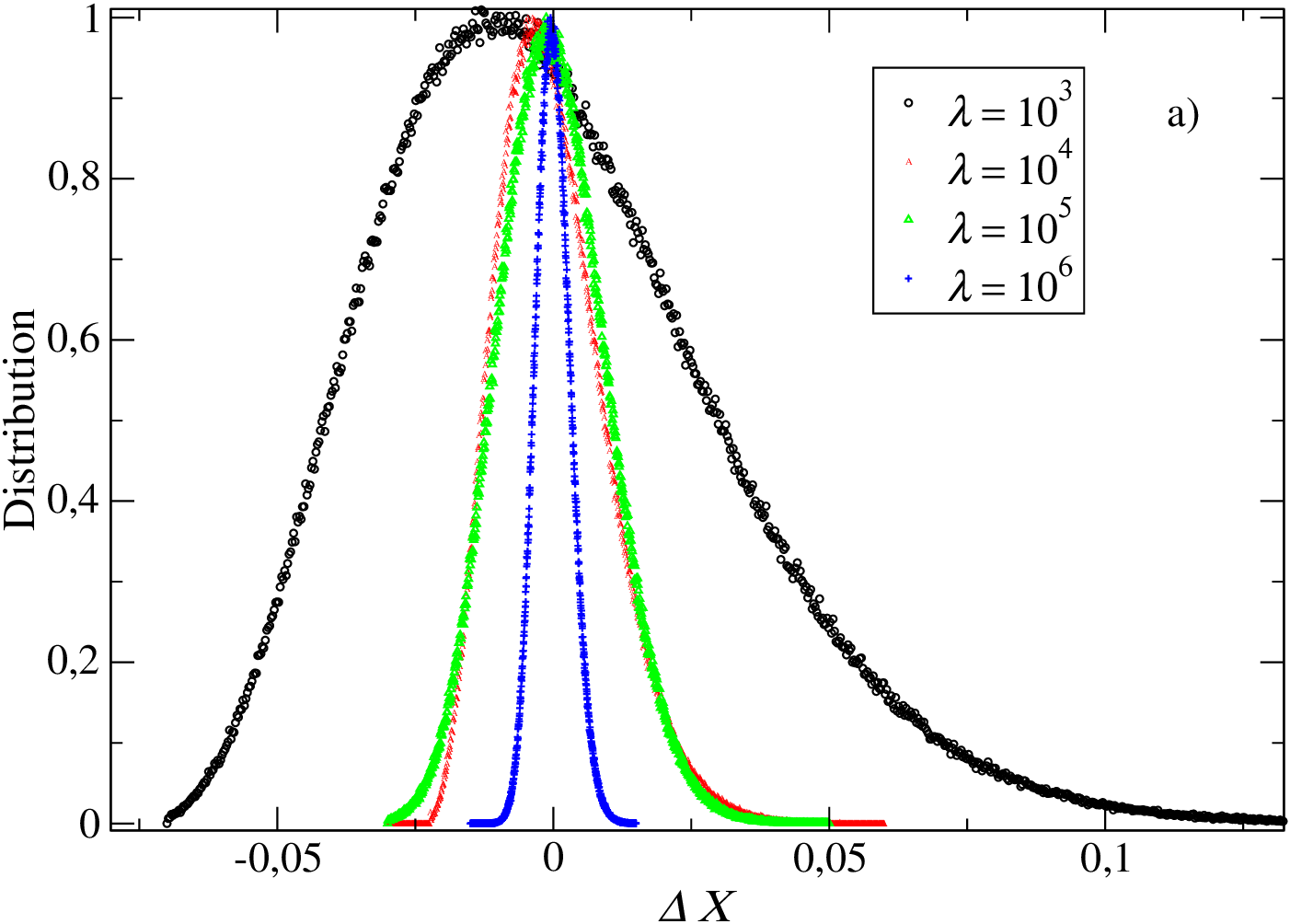}
    \includegraphics[width=8.1cm,height=6cm]{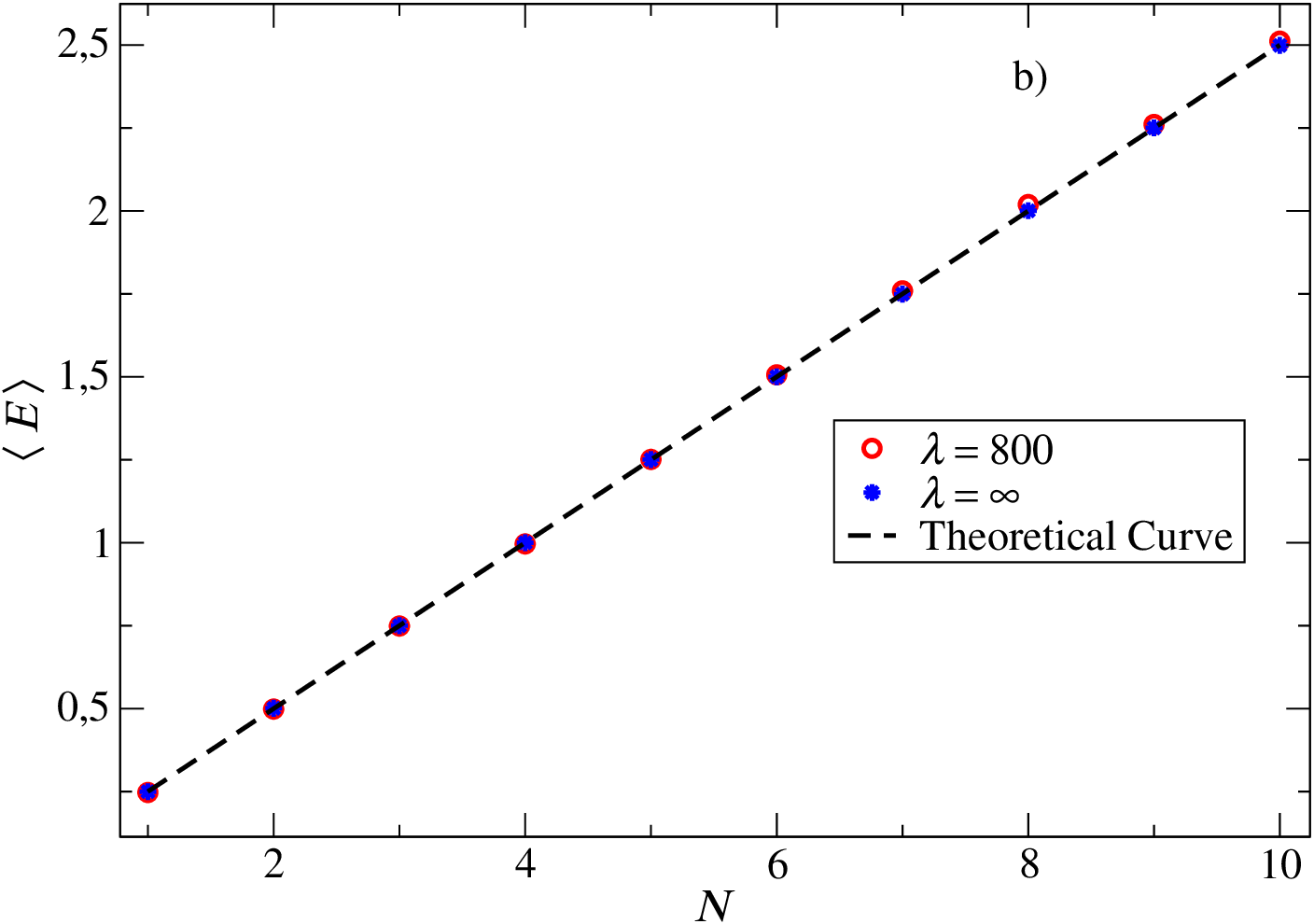}
    \caption{a) The distribution of the Compound Poisson Process  $\Delta X$ is calculated over $2*10^6$ integration points and plotted for various values of $\lambda$. For very large value of $\lambda$ $\Delta X$ tends to the Wiener process. b) The system's energy $\av{E}$ is calculated for $N=1,\ldots,10$ numbers of constituent molecules demonstrating its extensivity as well in an infinite heat bath $\lambda\to\infty$ as in a finite heat bath $\lambda=800$.}
    \label{Fig1}
\end{center}
\end{figure}

For our numerical simulations we set $\gamma=0.1$, $N=10$ and $\beta=2$ and rewrite the Langevin dynamics using the numerical solution's scheme in Ref. \cite{TalknerEtal} as
\begin{equation}
v(t_{i+1})=v(t_i)-\gamma v(t_i)\Delta t+\Delta X_i\,,
\end{equation}
where $\Delta X$ is the Compound Poisson Process. The former tends to the Wiener Process for $\mu\to \infty$ (see Ref. \cite{TalknerEtal} for details). This behaviour is demonstrated in Fig. \ref{Fig1}a) recording the distribution of
$\Delta X$ for four values of $\mu$. As we can see, by increasing $\mu$ the distribution become as expected more and more symmetric around zero approaching the Wiener process. Practically, we see that when $\mu$ is of the order of magnitude $10^6$  we are in the regime of an infinite heat bath. Accordingly, for lower orders of magnitude the heat is considered to be finite. In Fig. \ref{Fig1}b) we have plotted the mean energy of the system depending on the number of molecules $N=1,\ldots,10$, for $\lambda=800$, corresponding to a finite heat bath and for $\lambda\to\infty$ corresponding to an infinite heat bath. We can see that the extensivity property holds in both cases as predicted by our results.

\begin{figure}[t]
\begin{center}
    \includegraphics[width=8.1cm,height=6cm]{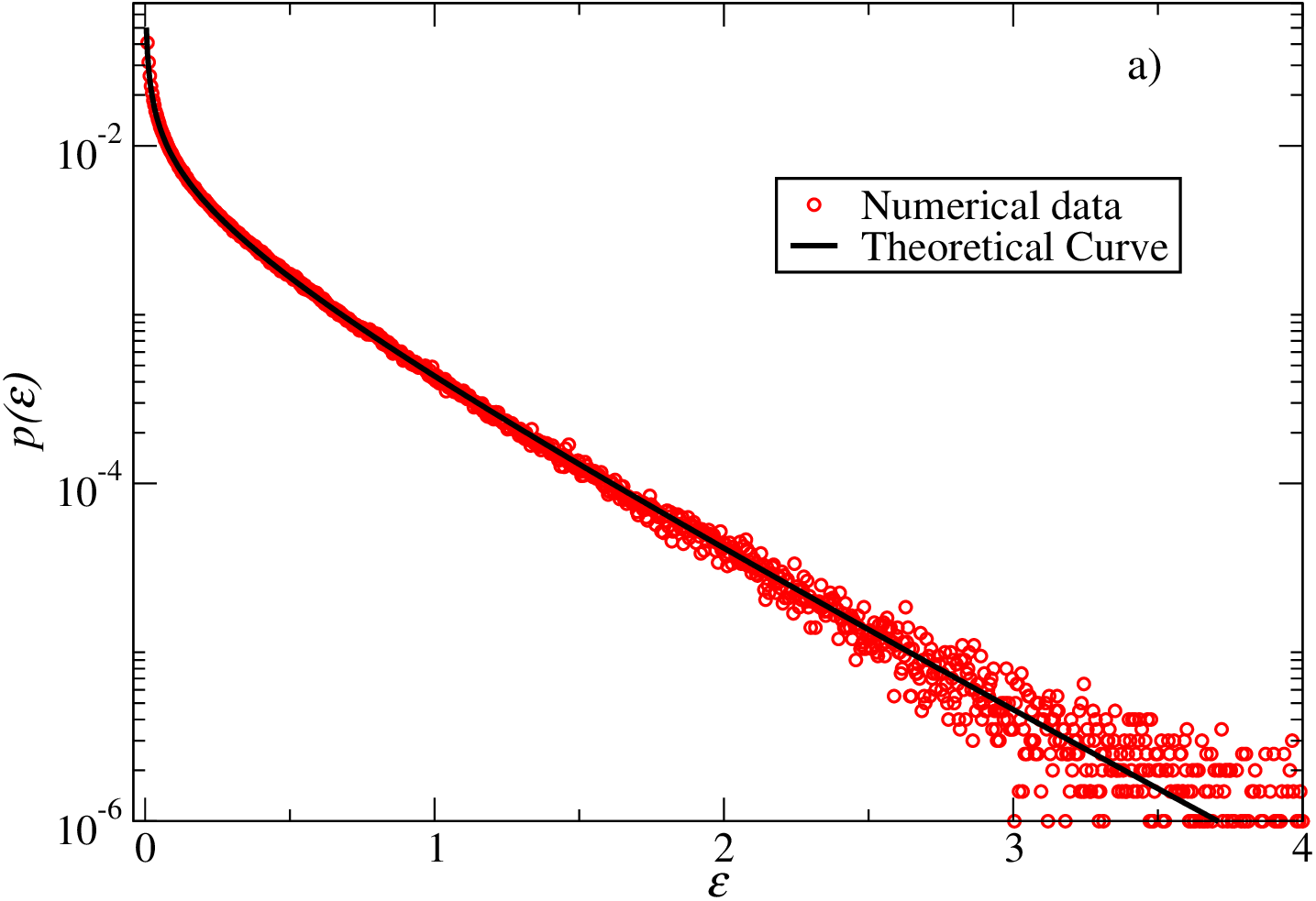}
    \includegraphics[width=8.1cm,height=6cm]{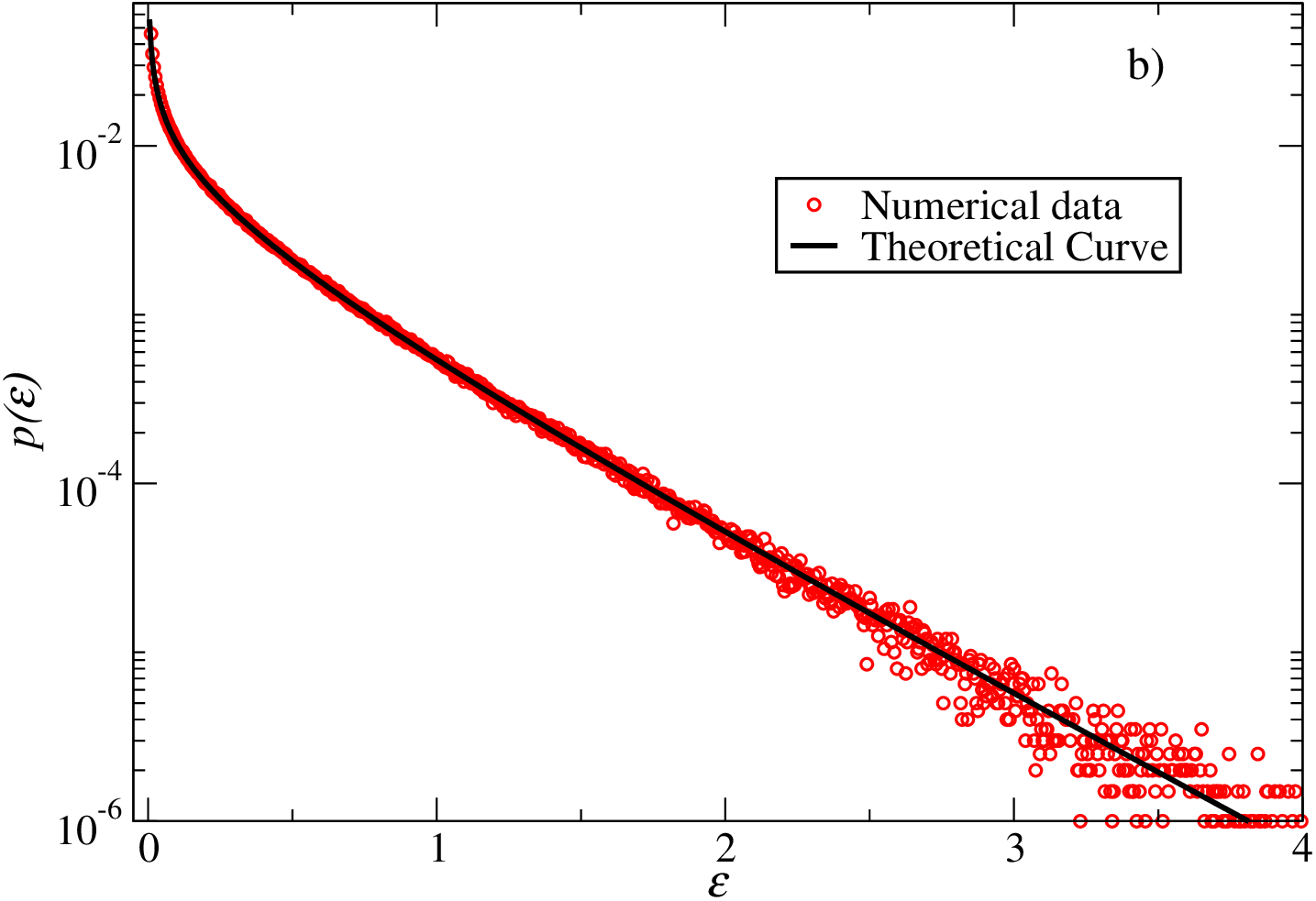}
    \caption{The free molecule continuous energy distributions of a single molecule (randomly chosen) a) in a finite heat bath $(\lambda=800)$ and b) in an infinite heat bath $(\lambda\to\infty)$ log-linear scale are recorded.
    The red circles represent the numerically obtained data over $2\times 10^6$ integration points and the solid black line is the theoretical curve in  Eq. (\ref{EnergyDistr_FreeMol}).}%
    \label{Fig2}
\end{center}
\end{figure}

\begin{figure}[t]
\begin{center}
    \includegraphics[width=8.1cm,height=6cm]{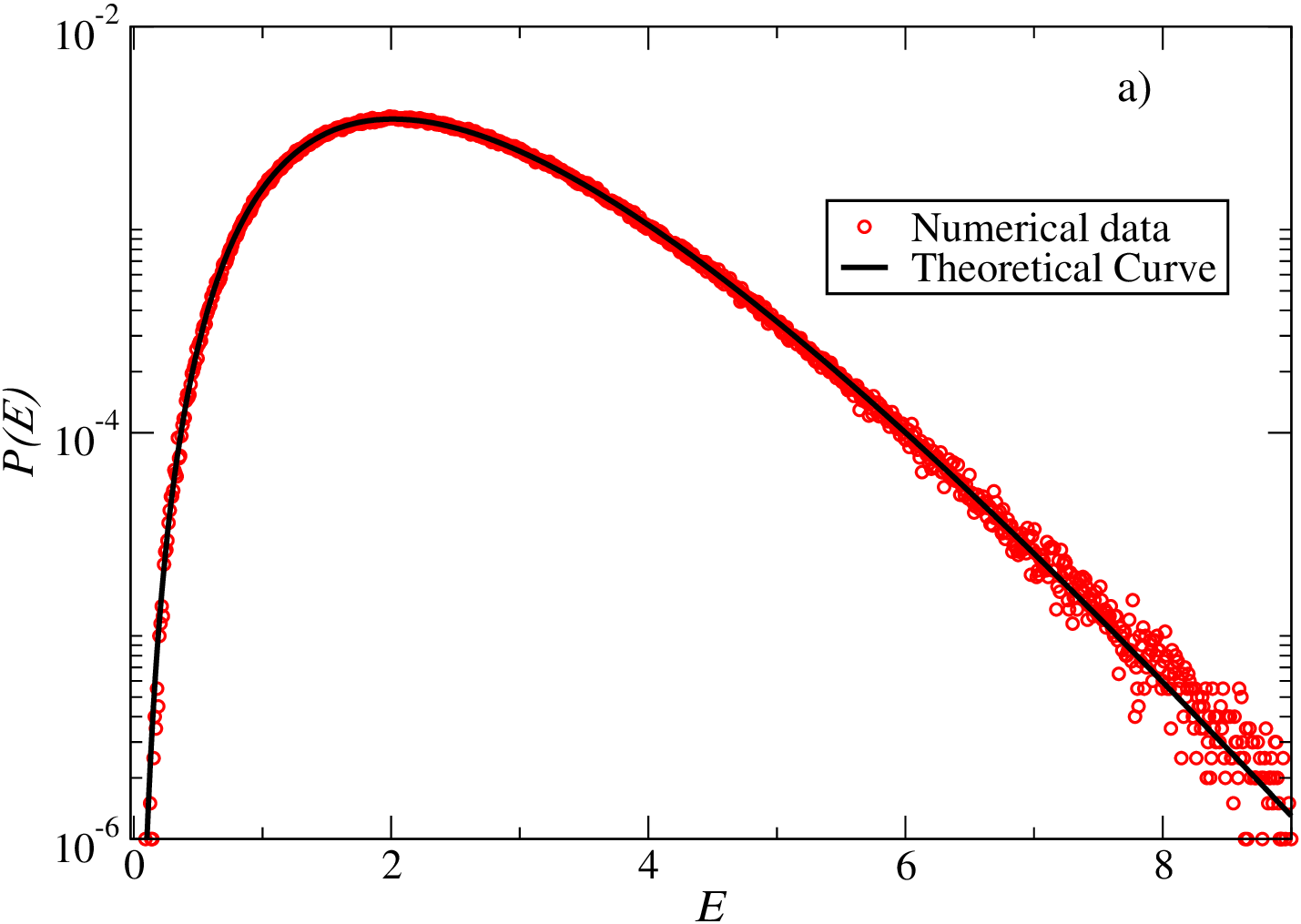}
    \includegraphics[width=8.1cm,height=6cm]{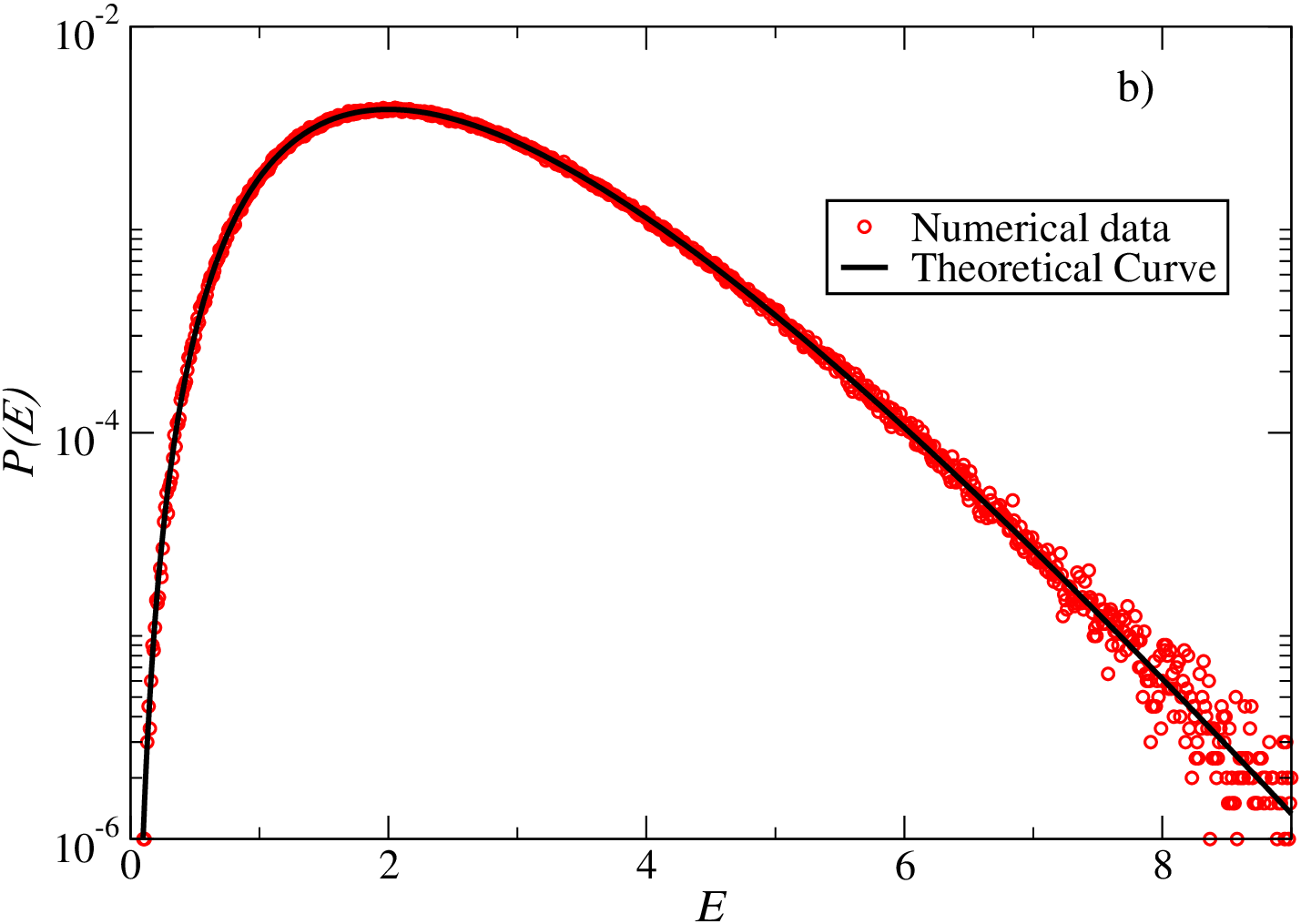}
    \caption{The free molecule continuous energy distributions of the total system of $N=10$ molecules a) in a finite heat bath $(\lambda=800)$ and b) in an infinite heat bath $(\lambda\to\infty)$ log-linear scale are recorded. The red circles represent the numerically obtained data over $2\times 10^6$ integration points and the solid black line is the theoretical curve in  Eq. (\ref{EnergyDistr_FreeMol}).}%
    \label{Fig3}
\end{center}
\end{figure}

The numerical energy distributions (red circles) and the
respective theoretical formula function (black solid line) in Eq. (\ref{EnergyDistr_FreeMol}) for a single
molecule in a finite ($\lambda=800$) and in an infinite heat bath are presented in Figs. \ref{Fig2}a) and \ref{Fig2}b), respectively.
As can be seen, the numerical results are in full agreement with the theoretical ones.
Similarly, in Figs. \ref{Fig3}a) and \ref{Fig3}b) we plot the energy distribution of the entire system of $N=10$ molecules,  for the preceding finite and infinite heat bath, respectively.
%


\section{Conclusions}\label{Conclusion}
%
Considering discreteness as the point of departure and relying
only on the multiplicative probabilistic composition rule and
additivity of the energy, we have shown that the extensivity of
the mean energy follows even for finite number of molecules.
Then, extending our analysis to the continuous case, we have
explicitly derived the canonical distribution without invoking the thermodynamic limit.
The derivation also shows that the usual assumption of the equal a priori probabilities is redundant for obtaining the canonical distribution.
We demonstrate numerically the emergence of the canonical distribution for systems composed of finite number of molecules exhibiting extensive mean energy behaviour using the one dimensional Langevin thermostat.

Finally, we note some differences between our work and the one by Khinchin \cite{Khinchin}: first, Khinchin makes use of equal a priori probabilities in order to obtain the canonical distribution whereas the present work only uses the statistical independence (see C1 and C2 above). Second, the canonical distribution is obtained only in the thermodynamic limit according to Khinchin while we have shown that the canonical distribution can be obtained without such a limit. In this sense, we have shown that the inverse power law distributions are not obtained as a result of the finiteness of the bath \cite{Campisi, BagciOik2013}.

Of particular interest in our approach is the function $\Phi(E)$ (or $\phi(\varepsilon)$), which can be essentially any arbitrary function. It is worth studying in a separate work whether and how its explicit structure depends on the number of molecules of the heat bath, or in other words, if its expression is determined from the heat bath - system interaction.

\section*{Appendix A: Proof of Eqs. (\ref{MC_relation_2})-(\ref{NewEq2_2})}\label{Appendix1}
%


In Eq. \eqref{ConfFun-ind-a}, we have determined the cardinality
$W$ of the sample space $\mathcal{W}$ by multiplying the
cardinalities of all the single molecules sample spaces
$\omega_\nu$.
A more detailed way of computing  $W$ is by means of the degeneracy number $\Omega(E_r)$ since by definition we must have $\sum_{r=1}^{w}\Omega(E_r) =\sum_{j=1}^{W}1 =W$. As explained in Section \ref{Ensemble}, $\Omega(E_r)$ of the ensemble $\mathcal{W}$ is given by the multinomial coefficient
\begin{equation}\label{MultiCoeff}
\Omega(E_r)=\frac{N!}{\prod_{i=1}^{\alpha}n_{ir}!}
=\prod_{i=1}^{\alpha-1}\binom{N-\sum_{k=1}^{i-1}n_{kr}}{n_{ir}}
\end{equation}
The summation over all $r$-energy states is equal to the summation of all frequencies, and thus
%
\begin{eqnarray}\label{multi-coef-ind}
\nonumber
\sum_{r=1}^{w}\Omega(E_r)&=&
\sum_{n_{1r}=0}^{N}\sum_{n_{2r}=0}^{N-n_1}\cdots\sum_{n_{\alpha-1,r}=0}^{N-\sum_{k=1}^{\alpha-2}n_{kr}}
\;\sqbr{\prod_{i=1}^{\alpha-1}\binom{N-\sum_{k=1}^{i-1}n_{kr}}{n_{ir}}}\\
&=&\sum_{n_{1r}=0}^{N}\binom{N}{n_{1r}}
\sum_{n_{2r}=0}^{N-n_1}\binom{N-n_{1r}}{n_{2r}}
\cdots
\sum_{n_{\alpha-1,r}=0}^{N-\sum_{k=1}^{\alpha-2}n_{kr}}
\binom{N-\sum_{k=1}^{\alpha-2}n_{kr}}{n_{\alpha-1,r}}=\alpha^N\,.
\end{eqnarray}
%
Here we have used  the relation $\sum_{i=1}^{\alpha}n_{ij}= N$ in Eq. (\ref{StateEnergy_j}).
Regarding now the normalization of the probabilities $P_j$ in Eq. (\ref{JointProb-jNotation}) it is fully equivalent to study it in terms of the normalization $P_r$ in Eq. (\ref{ESProbDF}).
Rewriting $P_r$ as follows
\begin{equation}\label{help-JP1}
P_r=\Omega(E_r)\prod_{i=1}^{\alpha}p_i^{n_{ir}}
=\Omega(E_r)p_\alpha^{N-\sum_{k=1}^{\alpha-1}n_{kr}}\prod_{i=1}^{\alpha-1}p_i^{n_{ir}}
=\Omega(E_r)p_\alpha^N\prod_{i=1}^{\alpha-1}\rbr{\frac{p_i}{p_\alpha}}^{n_{ir}}
=p_\alpha^N\prod_{i=1}^{\alpha-1}\binom{N-\sum_{k=1}^{i-1}n_{kr}}{n_{ir}}\rbr{\frac{p_i}{p_\alpha}}^{n_{ir}}\,,
\end{equation}
we obtain
%
\begin{eqnarray}\label{Norm1}
\sum_{j=1}^{W}P_j
= \sum_{r=1}^{w}P_r
=p_\alpha^N\sum_{n_{1r}=0}^{N}\sum_{n_{2r}=0}^{N-n_1}\cdots \sum_{n_{\alpha-1,r}=0}^{N-\sum_{k=1}^{\alpha-2}n_{kr}}\prod_{i=1}^{\alpha-1}\binom{N-\sum_{k=1}^{i-1}n_{kr}}{n_{ir}}\rbr{\frac{p_i}{p_\alpha}}^{n_{ir}}\,.
\end{eqnarray}
%
Then, from the Multinomial Theorem \cite{Proof1} we know the result of the r.h.s. of Eq. (\ref{Norm1}), namely
%
\begin{equation}\label{app1}
\sum_{j=1}^{W}P_j = \sum_{r=1}^{w}P_r
=\rbr{\sum_{i=1}^{\alpha}p_i}^N\,,
\end{equation}
%
which is exactly the result right above Eq.
(\ref{MC_relation_2}).
Moreover,  applying the operator $p_k \frac{\partial }{\partial
p_k}$ on Eq. (\ref{app1}), using the analytical expression of
$P_j$ in Eq. (\ref{JointProb-jNotation}), we obtain
%
\begin{align}\label{Proof-1b}
p_k\frac{\partial}{\partial p_k}
\sum_{j=1}^{W}P_j=
p_k\frac{\partial}{\partial p_k}\rbr{\sum_{i=1}^{\alpha} p_i}^N
\qquad\Longrightarrow\qquad
\sum_{j=1}^{W}P_j\; n_{ij}=
N\,p_i\rbr{\sum_{i=1}^{\alpha}p_i}^{N-1}\,,
\end{align}
%
which is exactly the result above Eq. (\ref{NewEq2_2}).
Taking  into account the normalization of the single molecule probabilities in Eq. (\ref{ConfFun-ind-a}), Eqs. (\ref{app1}) and \eqref{Proof-1b} yield  the results in Eqs. (\ref{MC_relation_2}) and (\ref{NewEq2_2}), respectively.

\section*{Appendix B: Convergence of Eq. (\ref{DiscExpr1b})}\label{Appendix2}
%
The energy value $E_r$ in Eq. (\ref{StateEnergy_j}) for
$j\rightarrow r$, can be rewritten as
\begin{equation}
E_r=\sum_{i=1}^{\alpha}n_{ir}\varepsilon_i=N \av{\varepsilon}_r\,,
\end{equation}
so that the difference between two successive energy states of the ensemble is equal to
\begin{equation}
\Delta E_r= N\sqbr{\av{\varepsilon}_{r+1} - \av{\varepsilon}_{r}}\,.
\end{equation}
Here we can read the following. The higher $\alpha$ is the more molecule energy states $\varepsilon_i$ we have with less distance between them, so that for great values of $\alpha$ we get
\begin{equation}\label{EnergyDiff}
\lim_{\alpha\rightarrow\infty}\Delta E_r \rightarrow
\mathrm{d}E\,.
\end{equation}

Next, we want to explore its convergence when $r\rightarrow\infty$ or equivalently $\alpha\rightarrow\infty$ (the latter is a necessary and sufficient condition to the former limit for finite $N$) of the discrete function $F(E_r)$ in Eq. (\ref{DiscExpr1b}). Considering now $F(E_{r})$  as sequence $F_r:=F(E_r)$ we shall use the following convergence criterion (ratio test),
\begin{equation}\label{convergence}
\text{If}\qquad
\lim_{r\rightarrow\infty}\abs{\frac{F_{r+1}}{F_r}}
=\lim_{\alpha\rightarrow\infty}\abs{\frac{F_{r+1}}{F_r}}<1
\qquad\text{then}\qquad F_r\rightarrow F\,.
\end{equation}
This criterion shows that the sequence $F_r$ converges to a value $F$ when $\alpha$ is taken into consideration. Then, we have
\begin{equation}
\lim_{\alpha\rightarrow\infty}\abs{\frac{F_{r+1}}{F_r}}
=\lim_{\alpha\rightarrow\infty}
\underbrace{\abs{\frac{\Omega(E_{r+1})}{\Omega(E_r)}}}_{=:G_1}
\underbrace{\abs{\frac{\Delta E^*_{r+1}}{\Delta E^*_r}}}_{=:G_2}
\underbrace{\abs{\exp\rbr{\sum_{i=1}^{\alpha-1}\Delta n_{i{r}}
\ln\big(f(\varepsilon_i)\Delta\varepsilon^*_i\big)}}}_{=:G_3}\,,
\end{equation}
where $\Delta n_{ir}:=n_{i,{r+1}}-n_{ir}$. By virtue of Eq.
(\ref{EnergyDiff}) we then have
\begin{equation}
\lim_{\alpha\rightarrow\infty}G_1=\lim_{\alpha\rightarrow\infty}G_2=1\,.
\end{equation}
Regarding the last  term $G_3$, we observe that the term $\Delta
n_{ir}$ takes values in the range $[0,N]$, namely finite values
for finite $N$.
On the other hand, assuming that $f(\varepsilon_i)$ converges to a continuous function, $f(\varepsilon_i)\Delta \varepsilon^*_i$ is less than unity  in the continuous limit, so that $\ln\big(f(\varepsilon_i)\Delta \varepsilon^*_i\big)\rightarrow-\infty$ and accordingly
\begin{equation}
\lim_{\alpha\rightarrow\infty}G_3=0\,.
\end{equation}
Therefore, the convergence criterion in Eq. (\ref{convergence}) is
satisfied for any energy value as long as the function $f$
converges.
%

\section*{Acknowledgments}
\noindent
GBB thanks the Nazarbayev University for funding his visit there.



\begin{thebibliography}{000}
\bibitem{Comment1} An example of the tensor product over the conjuction operator is presented. For $\alpha=2=N$ we have $\mathcal{W}=\omega_1
\otimes_{\wedge}\omega_2=\{\varepsilon_1,\varepsilon_2\}\otimes_{\wedge}
\{\varepsilon_1,\varepsilon_2\}=\{\varepsilon_1\wedge\{\varepsilon_1,\varepsilon_2\},
\varepsilon_2\wedge\{\varepsilon_1,\varepsilon_2\}\}=\{\varepsilon_1\wedge\varepsilon_1, \varepsilon_1\wedge\varepsilon_2, \varepsilon_2\wedge\varepsilon_1, \varepsilon_2\wedge\varepsilon_2\}$. For the sake of simplicity we denote the former four ($W=4$) configurations as $\{\mathcal{A}_1,\mathcal{A}_2,\mathcal{A}_3,\mathcal{A}_4\}$.
\bibitem{example} As an illustration of this subtle point, one can consider for example two independent rolls of a die so that the following three events may be observed: $A=\left\{ \text{1st roll is 1, 2, or 3}\right\}$, $B=\left\{ \text{1st roll is 3, 4, or 5}\right\}$, $C=\left\{ \text{the sum of the two rolls is 9}\right\}$. These events are not independent from one another as can be checked, e.g., from the relation $P\left( A\cap B\right) =\frac{1}{6}\neq \frac{1}{2}\frac{1}{2}=P\left( A\right) P\left( B\right)$. On the other hand, one can see that the multiplicative composition rule is still satisfied as can be seen from the equality $P\left( A\cap B\cap C\right) =\frac{1}{36} =\frac{1}{2}\frac{1}{2}\frac{4}{36}=P\left( A\right) P\left( B\right) P\left( C\right)$ (see Ref. \cite{Bertsekas2008} for more on this issue).
\end{thebibliography}
\end{document}